\documentclass[preprint,noshowpacs,noshowkeys,endfloats*]{revtex4}

\usepackage{dcolumn}
\usepackage{amsmath}
\usepackage{bm}
\usepackage{graphicx}
\usepackage{graphics}
\usepackage{longtable}
\usepackage{color}
\usepackage{epsfig}
\usepackage[normalem]{ulem}

\begin{document}

\title{Large magnetic gap at the Dirac point in a Mn-induced \BiTe\ heterostructure }

\author { E. D. L. Rienks$^{1,2,3,*}$,  S. Wimmer$^{4,*}$, P. S. Mandal$^{1,5}$, O. Caha$^6$, J. R\r{u}\v{z}i\v{c}ka$^6$, A. Ney$^4$, H. Steiner$^{4}$, V. V. Volobuev$^{4,7,8}$, H. Groiss$^9$,  M. Albu$^{10}$,  S. A.  Khan$^{11}$, J. Min\'ar$^{11}$, H. Ebert$^{12}$, G. Bauer$^4$, A. Varykhalov$^{1}$,  J. S\'anchez-Barriga$^{1}$,  O. Rader$^{1}$, G. Springholz$^4$ }  % end author
\affiliation{	$^1$Helmholtz-Zentrum Berlin f\"ur Materialien und Energie, 
			Elektronenspeicherring BESSY II, Albert-Einstein-Stra\ss e 15, 12489 Berlin, Germany}
\affiliation{$^2$  Institut f\"ur Festk\"orperphysik, Technische Universit\"at Dresden, 
			01062 Dresden, Germany  }
\affiliation{	$^3$   Leibniz-Institut f\"ur Festk\"orper- und Werkstoffforschung Dresden, 
			Helmholtzstra\ss e 20, 01069 Dresden, Germany}
\affiliation{$^4$Institut f\"ur Halbleiter- und Fest\"orperphysik, Johannes Kepler Universit\"at, 
			Altenberger Stra\ss e 69, 4040 Linz, Austria}
\affiliation{$^5$Institut f\"ur Physik und Astronomie, Universit\"at Potsdam, Karl-Liebknecht-Stra\ss e 
			 24/25, 14476 Potsdam, Germany}
\affiliation{$^6$  Department of Condensed Matter Physics, Masaryk University, 
			Kotl\'a\v rsk\'a 267/2, 61137 Brno, Czech Republic}
\affiliation{$^7$   National Technical University "Kharkiv Polytechnic Institute", Frunze Street 21, 
			61002 Kharkiv, Ukraine}
\affiliation{	$^8$   Institute of Physics, Polish Academy of Sciences, Al. Lotnikow 32/46, 
			02-668 Warsaw, Poland }

\affiliation{	$^9$ Zentrum f\"ur Oberfl\"achen- und Nanoanalytik, Johannes Kepler Universit\"at, Linz, 4040 Linz,
			 Austria}
\affiliation{$^{10}$  Graz Center for Electron Microscopy, Steyrergasse 17,  
			8010 Graz, Austria}
\affiliation{$^{11}$  New Technologies Research Centre, University of West Bohemia, Univerzitni 8, 
			306 14 Pilsen, Czech Republic}
\affiliation{$^{12}$  Department Chemie, Ludwig-Maximilians-Universit\"at M\"unchen, 
			Butenandtstr. 5-13, 81377 M\"unchen, Germany}
% end affiliation

\date{October 14, 2018}

\def\Tc{$T_{\rm C}$}
\def\BiTe{Bi$_2$Te$_3$}
\def\BiSe{Bi$_2$Se$_3$}
\def\SbTe{Sb$_2$Te$_3$}
\def\BiSbTe{(Bi, Sb)$_2$Te$_3$}
\def\SbVTe{(Sb$_{1-x}$V$_x$)$_2$Te$_3$}
\def\BiMnTe{(Bi$_{1-x}$Mn$_x$)$_2$Te$_3$}
\def\BiMnSe{(Bi$_{1-x}$Mn$_x$)$_2$Se$_3$}
\def\BiBiMnTe4{Bi$_2$MnTe$_4$}
\def\BiBiMnSe4{Bi$_2$MnSe$_4$}
\def\HgMnTe{Hg$_{1-x}$Mn$_x$Te}
\def\Gbar{\overline{$\Gamma$}}
\def\MnBi{Mn$_{\rm Bi}$} 
\def\muB{$\mu_{\rm B}$}

\begin{abstract}
{\bf Magnetically doped topological insulators enable the quantum anomalous Hall effect (QAHE) which provides quantized edge states for lossless charge transport applications
\cite{Onoda03,YuScience10,Chang13,CheckelskyNP14,KouPRL14,BestwickPRL15,Kandala2015,ChangCZNM15,GrauerPRB15}. 
The edge states are hosted by a magnetic energy gap at the Dirac point \cite{YuScience10} but all attempts to observe it directly have been unsuccessful. The size of this gap is considered  the clue to overcoming the present limitations of the QAHE, which so far occurs only at temperatures one to  two orders of magnitude below its principle limit set by the ferromagnetic Curie temperature \Tc\ \cite{ChangCZNM15,GrauerPRB15}. 
Here, we use low temperature photoelectron spectroscopy to unambiguously reveal the magnetic gap of Mn-doped \BiTe\ films which is present only below \Tc. Surprisingly, the gap turns out to be $\sim90$ meV wide, which not only exceeds $k_{\rm B}T$ at room temperature but is also 5 times larger than predicted by density functional theory \cite{HenkFliegerPRL12}. 
By an exhaustive multiscale structure characterization we show that this enhancement is due to a remarkable structure modification induced by Mn doping. Instead of a disordered impurity system, it  forms an alternating sequence of septuple and quintuple layer blocks, where  Mn is predominantly incorporated in the center of the septuple layers.
This self-organized heterostructure substantially enhances the wave-function overlap and the size of the magnetic gap at the Dirac point, as recently predicted \cite{Otrokov2DMat}. 
Mn-doped \BiSe\ forms a similar heterostructure, however, only a large, albeit nonmagnetic gap is formed. We explain both differences based on the higher spin-orbit interaction in \BiTe\ with the most important consequence of a magnetic anisotropy perpendicular to the films, whereas for \BiSe\ the spin-orbit interaction  it is too weak to overcome the dipole-dipole interaction.   
Our findings provide crucial insights for pushing the lossless transport properties of topological insulators towards room-temperature applications. }% end bf
 
\end{abstract}

\maketitle

The quantum anomalous Hall effect (QAHE) is characterized by a quantized Hall resistance  $\rho_{xy} = h/(Ne^2)$  where $N$ is an integer number $N$ of gapless 1D edge states and which does not require   the presence of an external magnetic field  \cite{Onoda03,YuScience10}. 
Magnetically doped 3D topological insulators of the tetradymite family \cite{YuScience10} have led to the first demonstration of the QAHE in Cr-doped \BiSbTe\  \cite{Chang13,CheckelskyNP14,KouPRL14,BestwickPRL15,Kandala2015}. 
Later on, the replacement of Cr by V as magnetic dopant delivered the first precise quantized values for $\rho_{xy}$ as well as a vanishing $\rho_{xx}$  at zero magnetic field \cite{ChangCZNM15,GrauerPRB15}, which is the key signature for lossless charge transport through edge channel devices \cite{ZhangSPIE12}. A crucial issue which would help to understand and develop the QAHE further towards applications has, however, remained fundamentally open -- the observation and quantification of the magnetic gap at the Dirac point \cite{YuScience10}. 

\textcolor{black}{In a magnetic topological insulator the QAHE occurs due to a modification of the band inversion, in which at the onset of ferromagnetic   order the inversion of one of the spin subbands is released by the exchange splitting and spin orbit coupling \cite{YuScience10}.}  
The observation of this  exchange splitting has, however, remained elusive. It manifests itself  as a magnetic gap at the Dirac point {\textcolor{black}{that opens when the system is cooled below the ferromagnetic transition temperature}}. The size of the gap is the crucial parameter for the temperature at which the QAHE can be observed. So far, this temperature is very low, typically around  $50$ mK \cite{GrauerPRB15}  to 2 K   \cite{MogiAPL15,Xiao2018}, which is one to two orders of magnitude lower than the ferromagnetic  \Tc\  of these systems. 
First principles calculations have recently suggested that the magnetic gap can be enhanced in topological insulator  heterostructures  \cite{Otrokov2DMat}. 

Angle resolved photoemission spectroscopy (ARPES)  is the method of choice for the direct observation of the magnetic gap and the verification of these predictions. Nevertheless, the experimental situation appears confusing: Large gaps at the Dirac point of the order of 0.1--0.2 eV of  \BiSe\ doped with Mn \cite{HasanPRB,JSB16} were explicitly shown to be {\em not} of magnetic origin \cite{JSB16}.
 {\textcolor{black}{On the contrary, no gaps are observed for  \BiSe\ when magnetic impurities are deposited directly on its surface  \cite{JSB16,VallaPRL12,ScholzPRL12,HonolkaPRL12}. Also, no gap appears when Mn is doped in the bulk of \BiTe\  where the Dirac cone was found to remain intact  at temperatures down to 12 K\ \cite{RCH15}. For V-doped \SbTe\ a mobility gap of 32 meV was inferred from scanning tunneling Landau level spectroscopy at  1.3 K  in comparison to pure \SbTe\ \cite{SessiNC16}, but due to the strong overlap with magnetic impurity states, no gap could be observed in the local density of states and no correlation to magnetism was reported.}}
 {\textcolor{black}{For Cr-doped \BiSbTe\ an average  gap of  56  meV was found by tunneling spectroscopy \cite{LeePNAS15}, but again its origin remained elusive because no temperature dependence was found.
In fact,  a gap as large as $\sim$75 meV was found for \BiSe\ with 4\%\ Cr by ARPES {\em{even at room temperature}} \cite{ChangCZARPESPRL14}. This clearly suggests a non-magnetic origin of these effects because the ferromagnetic \Tc\ is well below 50 K in all of these systems.}}

Interestingly, also the nature of the magnetic doping has   remained under debate. 
For isovalent doping, it was predicted that \BiSe, \BiTe\ and \SbTe\ will form a QAHE state. Thus, this should occur for doping with Cr or Fe but not for Ti or V  \cite{YuScience10}.  {\textcolor{black}{The fact that so far, V-doped (Sb,Bi)$_2$Te$_3$ displays the highest  QAHE  temperatures shows that the situation is more complex. In particular, non-isovalent magnetic dopants turn out to surprisingly little affect  the Fermi level and carrier concentration. For example, Mn-doped \BiSe\ and \BiTe\  always remain $n$-type \cite{RCH15,JSB16} despite the fact that divalent Mn replacing trivalent Bi should act as a strong acceptor. These puzzling issues are related to the comparatively complex tetradymite lattice structure consisting of quintuple layers separated by van der Waals gaps 
and the distribution of the dopants over a large number of electrically and magnetically inequivalent incorporation sites offered by the huge tetradymite unit cell consisting of thirty individual atoms. Hence, the actual magnetic dopant distribution is a key for understanding their impact on magnetism and band topology of the system. }}
 
To resolve these issues, we present here a comprehensive study of 
Mn-doped \BiTe\  and \BiSe\ films grown under nearly identical  conditions. Our data unequivocally reveals a pronounced magnetic exchange splitting at the Dirac point in \BiTe,  by far exceeding previous theoretical calculations \cite{HenkFliegerPRL12}. {\textcolor{black}{ It vanishes above the ferromagnetic Curie temperature, being clear cut evidence for its magnetic origin. On the other hand, no such gap could be identified for Mn-doped \BiSe\ even at 1 K.}
By multiscale characterization, we find that the actual lattice structure is  very different from the anticipated random impurity system:  Mn doping induces the formation of self-organized heterostructures consisting of stacked quintuple and septuple layers. This turns out to be very efficient for obtaining large magnetic gaps. As further crucial factors that distinguish the telluride and selenide systems, we identify the spin-orbit interaction and the way the heterostructures evolve  with the Mn concentration.

\phantom{xxxx}

\noindent {\bf Temperature-dependent band gap and magnetism}

\noindent
Figure  1 shows for Mn-doped \BiTe\ and \BiSe\  ARPES dispersions of the Dirac cone measured  above and below the ferromagnetic phase transition (\Tc\ = 10 and 6 K, respectively) {\textcolor{black}{for the same 6\% Mn concentration. For Mn-doped \BiTe, the photoemission spectrum from the center of the surface Brillouin zone shows an intensity maximum due to bulk transitions (50 eV photon energy), while the Dirac point ($E_{\rm D}$) of the surface state contributes a smaller peak at  $\sim$0.3 eV binding energy.  
Upon cooling of the sample from 20 K through \Tc\ down to 1 K, the low-binding energy flank of the peak develops a pronounced shoulder, forming a plateau around 0.2 eV, as can be seen in Fig. 1(a--c). We have quantified this change by linear fits to small sections of the spectrum, indicated in Fig. 1(c), with the condition that the slopes at 20 and 1 K are the same. The obtained shifts (S1 $\sim21$ meV $>$ S2 $\sim12$ meV) are compatible with the scenario in which a single component for the topological surface state at 20 K is split into two equally intense components at 1 K. A simulation of this scenario (Fig. 1(d)) shows that, for a reasonable parameter range, the gap that opens at the Dirac point is 2.5--3 times larger than the sum of shifts (S1 + S2). We therefore arrive at an estimate of the gap $\Delta$ of $90\pm10$ meV at low temperature.
Because  the ferromagnetic \Tc\ amounts to $\sim$10 K}} in this sample,  this is the clear and unambiguous proof of the magnetic origin of this gap. This is the central result of the present work. 

What is probed by the gap at the Dirac point is the exchange splitting of p-electrons of the  \BiTe\ host material which ferromagnetically couple  the localized magnetic moments of the Mn ions   \cite{HenkFliegerPRL12}. 
The  magnitude of the gap at the Dirac point depends on the exchange coupling $J$ and the magnetization along the [0001] surface normal    \cite{Rosenberg12}. This behavior is principally similar to that of the magnetic gap  of Mn-doped multiferroic GeTe   \cite{KrempaskyNC16} (although that system is of Rashba type and   topologically trivial). 
Returning to Fig. 1, we see that this criterion is {\it not}  fulfilled for Mn-doped  \BiSe\  with 6\% Mn, where we find a large gap at the Dirac point of $\sim$200 meV at all temperatures from 1 to 300 K \cite{JSB16}. The gap size does not increase when we cool down to 1 K, i.e., well below \Tc\ $\sim6$ K. This rules out a significant  contribution of magnetism to the Dirac cone gap in Mn-doped \BiSe,  contrary to the case of \BiTe,  {\textcolor{black}{and serves as an additional cross-check for the magnetic gap opening.}}

Figure 2 shows magnetization measurements for Mn-doped Bi$_2$Te$_3$ and Bi$_2$Se$_3$ samples with comparable Mn concentrations. The Bi$_2$Te$_3$ sample shows an easy axis magnetization {\bf M} along the $c$ axis normal to the surface, i.e., $M_{\perp} > M_{\|}$. This perpendicular anisotropy is robust since it does not depend on the Mn concentration (see Supplementary Information) and is observed also for bulk single crystals \cite{Hor10,Watson13}. In contrast, for Mn-doped Bi$_2$Se$_3$  the easy  axis is parallel to the surface plane -- also stable in a large concentration range (see Supplementary  Information). This   opposite behavior is also revealed by magnetotransport measurements shown in Fig. 2(c,d), where with magnetic fields applied perpendicular to the films  only Mn-doped \BiTe\ displays a pronounced anomalous Hall effect (AHE) upon cooling below  \Tc\, whereas it is negligible in \BiSe\ (Fig. 2(d)).

While an  in-plane magnetization merely shifts the Dirac cone in momentum space parallel to the surface \cite{Kharitonov,HenkFliegerPRL12}, the perpendicular anisotropy in Mn-doped \BiTe\ is  precisely the precondition for the magnetic band gap opening  and for the QAHE. 
Density functional theory (DFT) has  predicted the resulting magnetic Dirac gap as 16 meV for 10\% Mn substitutionally incorporated  in the topmost  quintuple layer  of \BiTe\  \cite{HenkFliegerPRL12}, but this gap has so far not been experimentally demonstrated. More astonishingly, the magnetic gap of $90\pm10$ meV that we observe is $5\times$ as large as the theoretical prediction.    

To resolve this discrepancy, we return to Fig. 2, where besides the obvious difference in magnetic anisotropy, there are a several other interesting differences between the two systems: Firstly, the coercive field of Mn-doped Bi$_2$Te$_3$ is significantly larger than for Bi$_2$Se$_3$, which only shows a very narrow opening of the hysteresis loop. Secondly, at the same time the anisotropy field, at which  in-plane and out-of-plane magnetizations are equal is two times higher for  Bi$_2$Te$_3$ (see Supplementary Information). Finally, the ferromagnetic $T_C$ of Mn-doped Bi$_2$Te$_3$ is considerably larger (7--15~K) than for Bi$_2$Se$_3$ (5--7~K) (see inserts in Fig. 2(a,b)) and depends more strongly on the Mn concentration. Altogether this demonstrates that Mn-doped Bi$_2$Te$_3$ is the more robust and anisotropic ferromagnet.

\phantom{xxxx}

\noindent {\bf Multiscale structure analysis}

\noindent
To clarify how Mn is actually incorporated in \BiTe\ and \BiSe\, a systematic multiscale structure analysis was performed for both types of samples.  
Figure 3(a) shows Mn-doped \BiTe\ in high-resolution scanning transmission electron microscopy (HRSTEM), measured along the $[\overline{1}100]$ direction for  10\%\ Mn. 
Strikingly, we observe upon Mn doping instead of the expected periodic sequence of Te-Bi-Te-Bi-Te quintuple layers    the emergence of a novel structure consisting of septuple and quintuple layers that does not exist for stoichiometric \BiTe.
The septuple layers consist of the sequence Te-Bi-Te-Mn-Te-Bi-Te, where the Mn atoms occupy the central septuple atomic layer as found in \BiBiMnTe4\ crystals \cite{DSLee13}. This self-organized   heterostructure formation obviously disagrees with the common notion of substitutional Mn incorporation in \BiTe\ assumed in most previous studies \cite{HenkFliegerPRL12,Schmidt11,Hor10}.  

Mn is not isoeletronic to Bi due to the different number of valence electrons. Therefore, substitutional Mn on Bi sites should be a strong acceptor inducing a strong p-type doping of the system. This is neither observed for Mn-doped \BiTe\ nor  \BiSe\ which always remain n-type even at high Mn concentrations  \cite{Watson13,Lee2014,RCH15,JSB16}. On the other hand, Mn incorporated in \BiBiMnTe4\ septuple layers is not electrically active because the septuple is formed by addition of a charge compensated MnTe double layer to a quintuple layer. Apart from compensation effects, this explains the surprisingly small effect of Mn-doping on  carrier concentration and Fermi level of the system.  
According to Fig. 3(b),  the formation of Mn-induced septuple/quintuple heterostructures also occurs in \BiSe\ and the  formation of septuple layers was also recently seen for Mn-doped \BiSe\  films \cite{HagmannNJP17}.

To obtain element specific information on the Mn incorporation sites, x-ray absorption near-edge spectroscopy (XANES) and extended x-ray absorption fine structure spectroscopy (EXAFS) were performed at the Mn$K$-edge  as summarized in Fig. 4. The absorption spectra were analyzed by simulations 
for all possible Mn incorporation sites, ranging from Mn in the center of the Bi$_2$MnTe$_4$ (Bi$_2$MnSe$_4$) septuple layers, substitutional Mn on Bi sites in the quintuple layers, interstitial Mn in the 
van der Waals gap in either octahedral or tetrahedral coordination (see Fig. 4(e--g)). In addition, we considered also Mn on anion Te (Se) antisites 
i.e., on Te1 (and Se1) sites at the outer layers and on Te2 (and Se2) sites in the center of the quintuple layers. Contrary to most theoretical investigations based on substitutional Mn incorporation \cite{HenkFliegerPRL12}, our analysis shows that Mn in \BiTe\ prefers to be incorporated in   septuple layers {\textcolor{black}{and only a minority as substitutional Mn in the quintuple layers. While the EXAFS data does not rule out Mn on octrahedral  sites in the van der Waals gap, the fact that septuples are never seen in undoped \BiTe\ clearly suggests that the Mn sites are closely linked to the septuple layers.}} 
 
Turning to Mn-doped  \BiSe\ shown in Figs. 4(b,d), we do not observe as intense EXAFS oscillations as for \BiTe. This indicates stronger cancellation effects caused by Mn distributed over different lattice sites, including a larger amount of substitutional Mn and a lesser fraction within the septuple layers.  
This is highlighted by the XANES spectra at the Mn$K$-edge which exhibit a characteristic double peak structure at higher and lower photon energy, attributed respectively to Mn in the center of the septuple and to substitutional Mn in the quintuple layers, where again for \BiSe\ the signal from Mn in the septuple layers is weaker as compared to Mn in \BiTe. 
 For tetrahedrally coordinated interstitial Mn, as well as for Mn on Te (Se) antisites, the simulations do not agree with the experiments, indicating that these are not favorable for Mn incorporation. Overall, we conclude that for \BiTe\ the vast majority of  Mn is incorporated within the septuple layers and that substitutional Mn is more readily formed in Bi$_2$Se$_3$, especially at lower Mn concentrations, with an overall broader distribution of Mn over various other lattice sites.   
 
Our results so far suggest a unique heterostructure formation upon Mn-doping but the HRSTEM, XANES, and EXAFS data deliver only a local picture. To systematically characterize its evolution on a larger length scale and its dependence on the Mn concentration,  x-ray diffraction investigations were performed as summarized in Fig. 5. For both systems we indeed find a pronounced change of the diffraction spectra with increasing Mn content. This is evidenced by the appearance of additional diffraction peaks (see Fig. 5(a,b)) that signify the emergence of septuple layers in the structure. However, the substantial broadening of the peaks reveal  that the septuple layers are not incorporated periodically at fixed distances, but rather stochastically after a varying number $N_{QL}$ of quintuple layers as seen in the STEM cross-sections where $N_{QL}$ varies between one to seven. This requires us to develop a one-dimensional paracrystal model to describe and evaluate the experimental diffraction data, as detailed in the Supplementary Information. In this model, the overall structure is described as a statistically varying sequence of quintuple segments alternating with single septuple layers that is characterized by the average number $\left< N_{\rm QL} \right>$  between subsequent septuples and by the randomness of the statistical distribution of the  $N_{QL}$, i.e., their root mean square (RMS) deviation from the average value. 

The model fits displayed by the black lines in  Figs. 5(a,b) show a remarkably good agreement with the diffraction spectra for all Mn concentrations. This impressively corroborates the formation of self-organized  quintuple/septuple layer heterostructures in both the \BiTe\ and \BiSe\ systems. From the fits, the average $\left< N_{\rm QL} \right>$ between the septuple layers
(and, thus, the density of septuple layers) as a function of Mn content is obtained, which is displayed in Fig. 5(c) together with the statistical variation of $N_{QL}$. Apparently, in \BiTe\ the formation of septuple layers starts at lower Mn concentration and the average separation between the septuple layers is substantially smaller than in \BiSe . This evidences the higher probability of septuple layer formation for \BiTe , in agreement with the XANES and EXAFS result. This difference is further highlighted by Fig. 5(d), where the derived number of available Mn sites in the septuples is plotted versus the actual Mn concentration, revealing that in \BiTe\ all Mn atoms can be incorporated in the septuple layers, whereas in \BiSe\ the density of septuple layers  at low  Mn concentrations is too small to accommodate all Mn atoms, i.e., a significant Mn fraction must be incorporated on other sites as well.   

\phantom{xxxx}

\noindent {\bf Discussion}

\noindent
The electronic structure of transition metal impurities at the surface and in the bulk of \BiTe\ and \BiSe\ has been studied extensively by density functional theory (DFT) calculations. A magnetic band gap of  $\sim$10 meV has  been predicted for Co-doped \BiSe\ \cite{Schmidt11} and of 16 meV  for  Mn-doped \BiTe\  \cite{HenkFliegerPRL12}. 
By the splitting of the Dirac point we probe the exchange interaction at the Te sites. This has been confirmed by  DFT, where the spectral density of the split Dirac point is nine times more strongly localized at the Te atoms than at the other sites \cite{HenkFliegerPRL12}. The calculated gap value of 16 meV for substitutionally Mn-doped \BiTe\ by Henk et al. \cite{HenkFliegerPRL12} for a Mn concentration of 10\%\  is, however, substantially smaller that the magnetic gap of 90 meV revealed by our experiments.
For  Mn in \BiSe, a nonmagnetic band gap of the measured size ($\sim$200 meV) does not appear in any DFT calculation. At least, DFT reveals in principle that, depending on orbital symmetry, small gaps in the Dirac cone may open due to hybridization with transition metal impurity states. Accordingly, a hybridization gap of $\sim4$ meV at the Dirac point was suggested for substitutional Mn in \BiSe\    for an in-plane ({\it sic!})  magnetization \cite{Abdalla13}, which is obviously much less than what we experimentally observe. The only prediction of a  such a nonmagnetic gap which also has the  magnitude seen in our experiments  is from calculations assuming an on-site Coulomb interaction $U$ at the impurity site \cite{Black1}.  

The structural information gained in  the present study helps  to clarify the question of the origin of this nonmagnetic gap: Apparently, Mn forms more substitutional sites in \BiSe\ than in \BiTe\ where Mn is preferentially incorporated into septuple layers.  
Mn in the substitutional site will lead to a larger Coulomb $U$ than in the central Mn monolayer of the septuple layer, where Mn $3d$ levels can delocalize in the plane. We could recently demonstrate experimentally  that $U$, termed previously as impurity strength \cite{Black1}, indeed influences the size of the nonmagnetic gap strongly. 
For example, indium impurities are known to enhance the spacing across which surface-surface coupling opens a gap at the Dirac point    \cite{Armitage}. 
When we compare  Mn doping with In doping in \BiSe, we find that to reach the same gap size as for 8\%\ Mn, only 2\%\ In is required \cite{JSB16,indium}. However, the effect of impurities on the nonmagnetic gap decreases with higher spin-orbit interaction of the host material  as shown by our recent work \cite{indium}. Thus,  \BiTe\ is principally less susceptible to opening of a nonmagnetic gap than \BiSe, regardless of the structural differences induced by  Mn doping. 

To explain the marked difference in the magnetic anisotropy of  Mn-doped \BiTe\  and \BiSe\  including the structural motif of the septuple layers, we compute the magnetic anisotropy using ab initio calculations (see Supplementary Information) for unit cells consisting of one Bi$_2$MnX$_4$ septuple layer and one Bi$_2$X$_3$  quintuple layer (X=Te, Se). We find that the magnetocrystalline anisotropy inducing the out-of-plane spin orientation is 3.5 times larger in the telluride than in the selenide system. This is due to the higher spin-orbit interaction in \BiTe\ and is related to warping effects in the Dirac cone, which turn the spins out of the plane \cite{Fu09}. The magnetocrystalline anisotropy is counteracted by the dipole-dipole interaction  (shape anisotropy) that tends to  align the magnetic moments in the plane. The shape anisotropy comes out to be very similar for the two heterostructure systems, but  essentially cancels the magnetocrystalline anisotropy in \BiSe, whereas it is superseded by the magnetocrystalline anisotropy in \BiTe. Thus, the out-of-plane anisotropy persists and is nearly one order of magnitude larger than for the selenide system, where the anisotropy energy almost changes sign towards in-plane magnetization. In the real samples, the in-plane magnetization for the selenide structures may be additionally supported by the Mn atoms on substitutional sites which favor the magnetic moments to be in plane \cite{Abdalla13}.  For the telluride system, the preferred perpendicular magnetization agrees well with recent model calculations by Otrokov et al. \cite{Otrokov2DMat} who considered similar types of \BiBiMnTe4/\BiTe\ heterostructures in different septuple and quintuple combinations. We conclude that the higher spin-orbit interaction in the telluride system thus  overcomes the dipole-dipole interaction and enables the formation of the magnetic gap at the Dirac point.

Finally, as mentioned above, our measured magnetic gap size of 90 meV for Mn-doped \BiTe\ is five times as large as predicted for substitutional Mn incorporation (16 meV) \cite{HenkFliegerPRL12}. This represents a huge enhancement that is obviously related to the naturally formed quintuple/septuple layer heterostructures. 
As pointed out by Otrokov et al. \cite{Otrokov2DMat} the Mn monolayer in the center of \BiBiMnTe4\ septuples  enhances the wave function overlap strongly, supporting magnetic gaps as high as 38 -- 87 meV, depending on the chosen \BiBiMnTe4/\BiTe\ combination. This is in excellent agreement with the enhancement found in our experiments. This demonstrates the great potential of such structures for stabilizing edge transport in QAHE devices. Theory \cite{Otrokov2DMat} also suggest that the nontrivial topology is retained, by the calculation of the Chern number   $C=-1$  for the heterostructure system, consisting of one \BiBiMnTe4 septuple and two \BiTe\  quintuple  layers, and by the persistence of the Dirac cone surface state.  This is confirmed by our ARPES measurements above and below \Tc.
 
In conclusion, we have demonstrated unambiguously and for the first time the opening of a magnetic gap in a topological insulator below the ferromagnetic phase transition and its closure for $T >T_C$. The magnetic gap in Mn-doped \BiTe\ is remarkably large (90$\pm$10 meV) as a result of the  formation of a natural heterostructure in which Mn is incorporated within septuple layers instead of simple substitutional incorporation. Our results thus support recent theoretical predictions that magnetic gaps in topological insulators can be significantly enhanced in multi-layered systems \cite{Otrokov2DMat}. These  are considered as a basis for the realization of new topological phases such as the axion insulator state exhibiting quantized magnetoelectric effects \cite{Xiao2018,Mogi2017} and the chiral Majorana fermion \cite{HeQLScience17}.
No magnetic gap is detected for Mn-doped \BiSe\ within the experimental resolution but instead, a very large non-magnetic gap that does not increase even when cooling down to 1 K, well below $T_C$, and thus does not have a magnetic contribution. We correlate this  with the difference in magnetic anisotropy due to the much larger spin-orbit interaction in \BiTe\ and offer a unifed picture for both observations.
Returning to the question of enhanced QAHE devices,  up to now the focus has been on increasing the Curie temperatures of the systems, {\textcolor{black}{e.g., by obtaining a high density of st
ates at the Fermi level to increase the exchange integrals.
Instead, the magnetic gap size may turn out to be the more decisive factor for pushing up the operation temperature.   
Due to the large magnetic gap size, Mn-doping seems to be most promising in this respect and will open up new perspectives for device realization.}}

\phantom{xxxx}

\noindent {\bf Data availability}

\noindent The data sets generated and analysed during the current study are available from the corresponding authors on reasonable request.

\phantom{xxxx}

\noindent {\bf Code availability}

\noindent The code for the paracrystal model is available from the
  corresponding authors upon request. The employed electronic structure codes Wien2K and 
  SPR-KKR and x-ray absorption fine structure codes FDMNES and FEFF9 can 
   be downloaded after the corresponding licence requirements given on the 
  respective webpages are fulfilled. 

\phantom{xxxx}

\noindent {\bf Acknowledgements}

\noindent {We thank B. Henne, F. Wilhelm, and A. Rogalev for support of the XANES and EXAFS measurements at ID 12 and BM23 beam lines of the ESRF, V. Hol\'y for advices on  the structure model,  W. Grafeneder for the TEM sample preparation and G. Bihlmayer and A. Ernst for helpful discussions. 
  S.A.K and J.M. are grateful for support from CEDAMNF  (CZ.02.1.01/0.0/0.0/15$\underline{\ }$003/0000358) of Czech ministry MSMT.
This work was partially supported by CEITEC Nano RI (MEYS CR, 2016–2019), by   SPP1666 of
Deutsche Forschungsgemeinschaft, and by Impuls- und Vernetzungsfonds der Helmholtz-Gemeinschaft
(Helmholtz-Russia Joint Research Group HRJRG-408 and Helmholtz Virtual Institute ``New states of matter and their excitations").  }

\phantom{xxxx}
\noindent {Corresponding authors: } 
G. Springholz, email: gunther.springholz@jku.at, O. Rader, email: rader@helmholtz-berlin.de.

$^*$ These authors contributed equally to the present work.

\begin{figure}[h]
  \begin{center}
    \includegraphics[width=\linewidth,keepaspectratio]{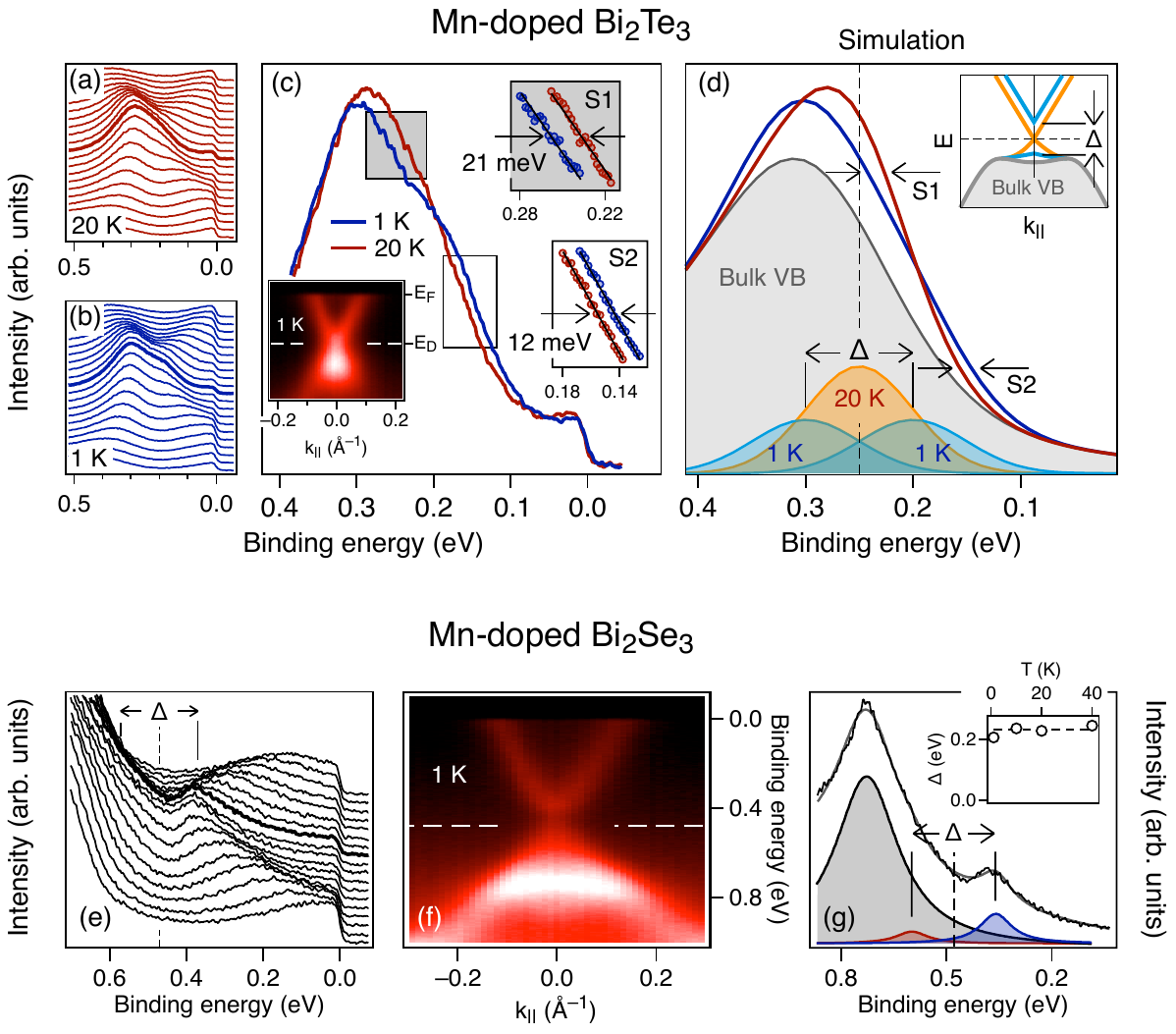}
	\vspace{1cm}
    \caption{ {\bf{Magnetic gap of Mn-doped  \BiTe\  derived by ARPES. }} (a--d) Measurements  for \BiTe\ with 6\% Mn   performed  above and below the Curie temperature \Tc\ $\sim10$ K. (The spectra in (c,d,g) and those marked by thick lines in (a,b,d) correspond to  the center of the surface Brillouin zone, i. e., electron wave vector component $k_\parallel=0$ \AA$^{-1}$.)
Linear fits to the regions indicated in (c) yield shifts of 21 and 12 meV between these sections of the 20 K and 1 K spectra.
(d) Simulation showing that this corresponds to a magnetic gap $\Delta=90\pm10$ meV. (e--g) Same analysis for Mn-doped  \BiSe\ with 6\% Mn and a \Tc\ of 6 K, revealing only a nonmagnetic gap of $220\pm5$ meV at 20 K and $205\pm5$ meV at 1 K, determined by  least-square fit  to the upper Dirac cone and to    the lower Dirac cone at $k_\parallel=0$ \AA$^{-1}$.   
}
  \end{center}
\end{figure}

\begin{figure}[t]
  \begin{center}
    \includegraphics[width=\linewidth,keepaspectratio]{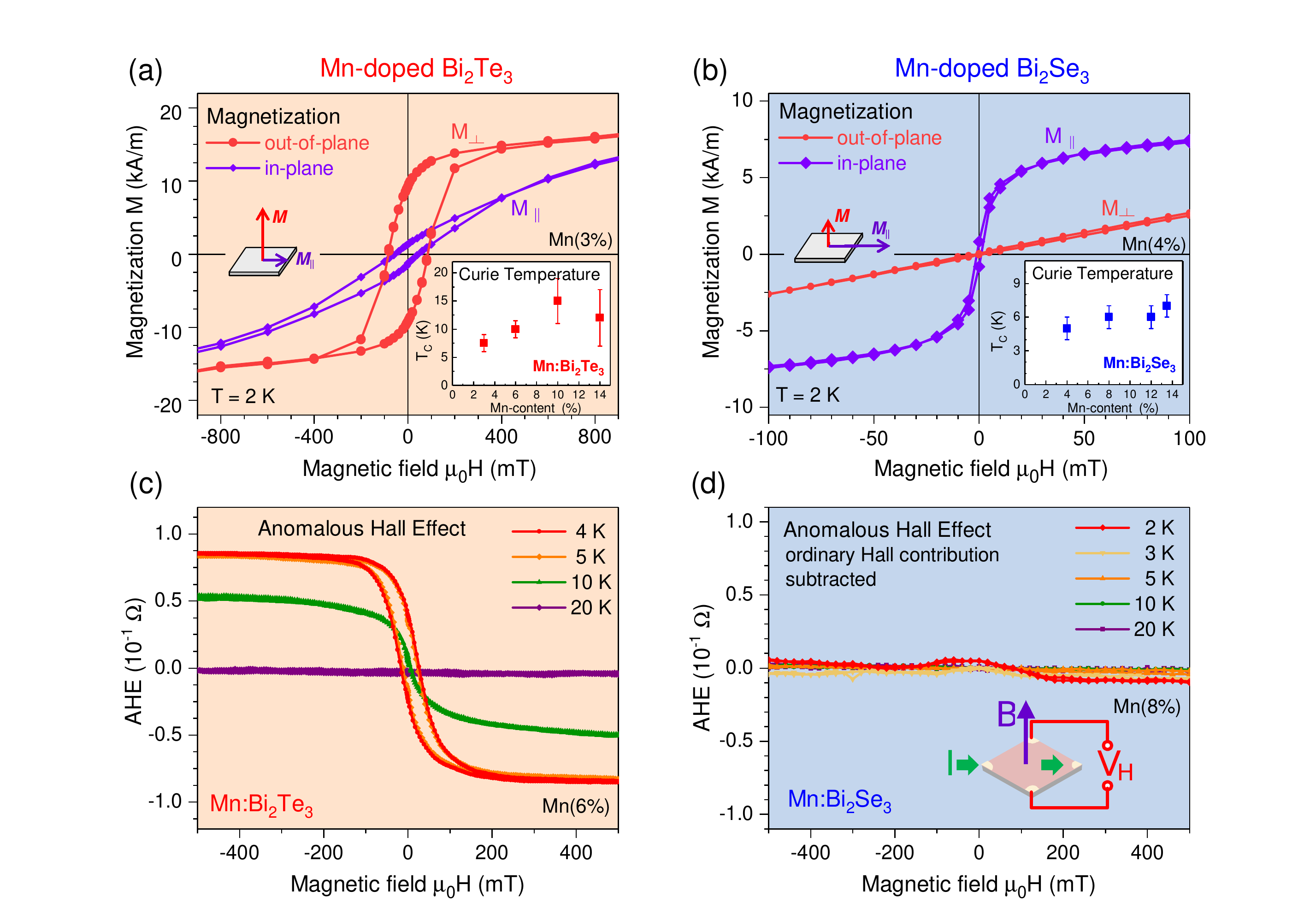}
	\vspace{1cm}
    \caption{{\bf{Magnetic properties.}} In-plane and out-of-plane magnetization $M(H)$ of \BiTe\  (a) and \BiSe\ (b) films with Mn concentrations of 3 and 4 \% measured at 2 K by SQUID with the magnetic field either parallel or perpendicular to the surface, evidencing a perpendicular anisotropy (easy axis) for \BiTe\ and an in-plane easy axis for \BiSe . The Curie temperature as a function of Mn concentration is depicted in the inserts, evidencing that \Tc\  is significantly higher in the telluride system. (c,d) Anomalous Hall effect (AHE) measurements of the samples with the contribution of the ordinary Hall effect extracted from the high field data subtracted (see Supplementary Information). Due to the perpendicular magnetic anisotropy, only Mn-doped \BiTe\ displays a pronounced anomalous Hall effect appearing when the sample is cooled below \Tc . 
}
  \end{center}
\end{figure}

\begin{figure}[t]
  \begin{center}
    \includegraphics[width=12cm,keepaspectratio]{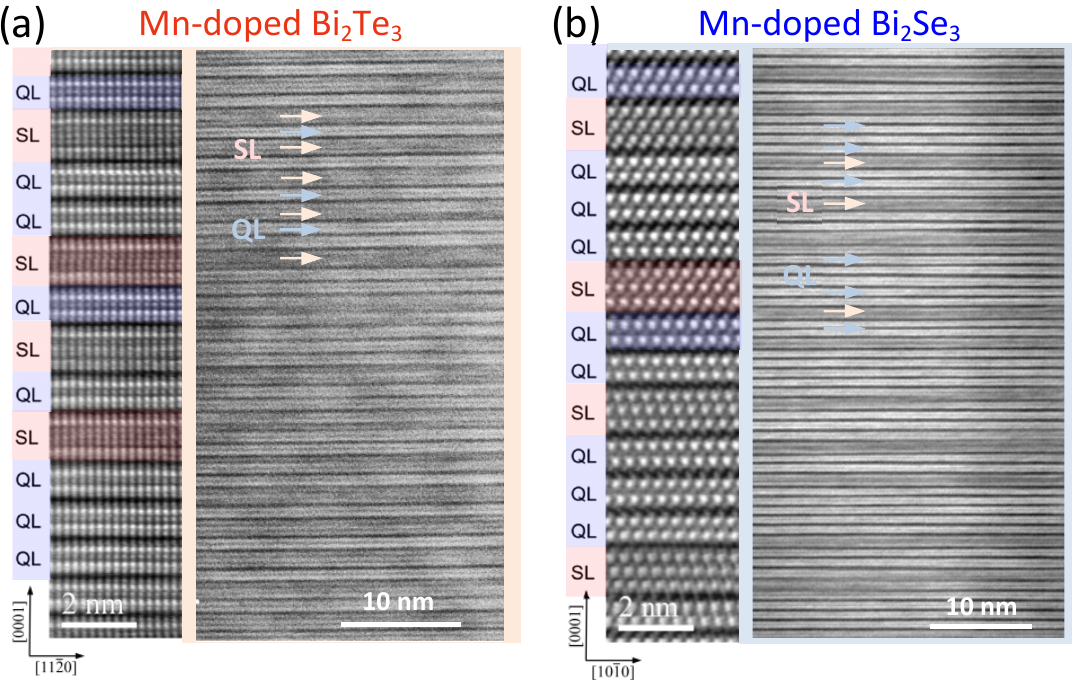}
	\vspace{1cm}
    \caption{\textcolor{black}{{\bf{HR-STEM cross sections}} of Mn-doped \BiTe\ and \BiSe\   recorded along the [$\overline{1}$100] and [$\overline{1}$2$\overline{1}$0] zone axis, respectively.}  The STEM cross sections reveal the natural formation of a layered heterostructure consisting of Bi$_2$MnTe$_4$ (Bi$_2$MnSe$_4$) septuple  layers (SL) inserted between Bi$_2$Te$_3$ (Bi$_2$Se$_3$) quintuple (QL) layers adjoined by  van der Waals gaps. Due to the atomic-number contrast, the heavy atoms (Bi) appear brighter in the high angle annular dark field (HAADF) images, and the septuple layers in the overview images darker due to the lighter Mn atoms preferentially incorporated. Note the different scales. The Mn concentration in (a) was 10\%\ and in (b) locally 9\%\ and on average 6\%\ according to x-ray diffraction. 
}
  \end{center}
\end{figure}

\begin{figure}[h]
  \begin{center}
	 \includegraphics[width=\linewidth,keepaspectratio]{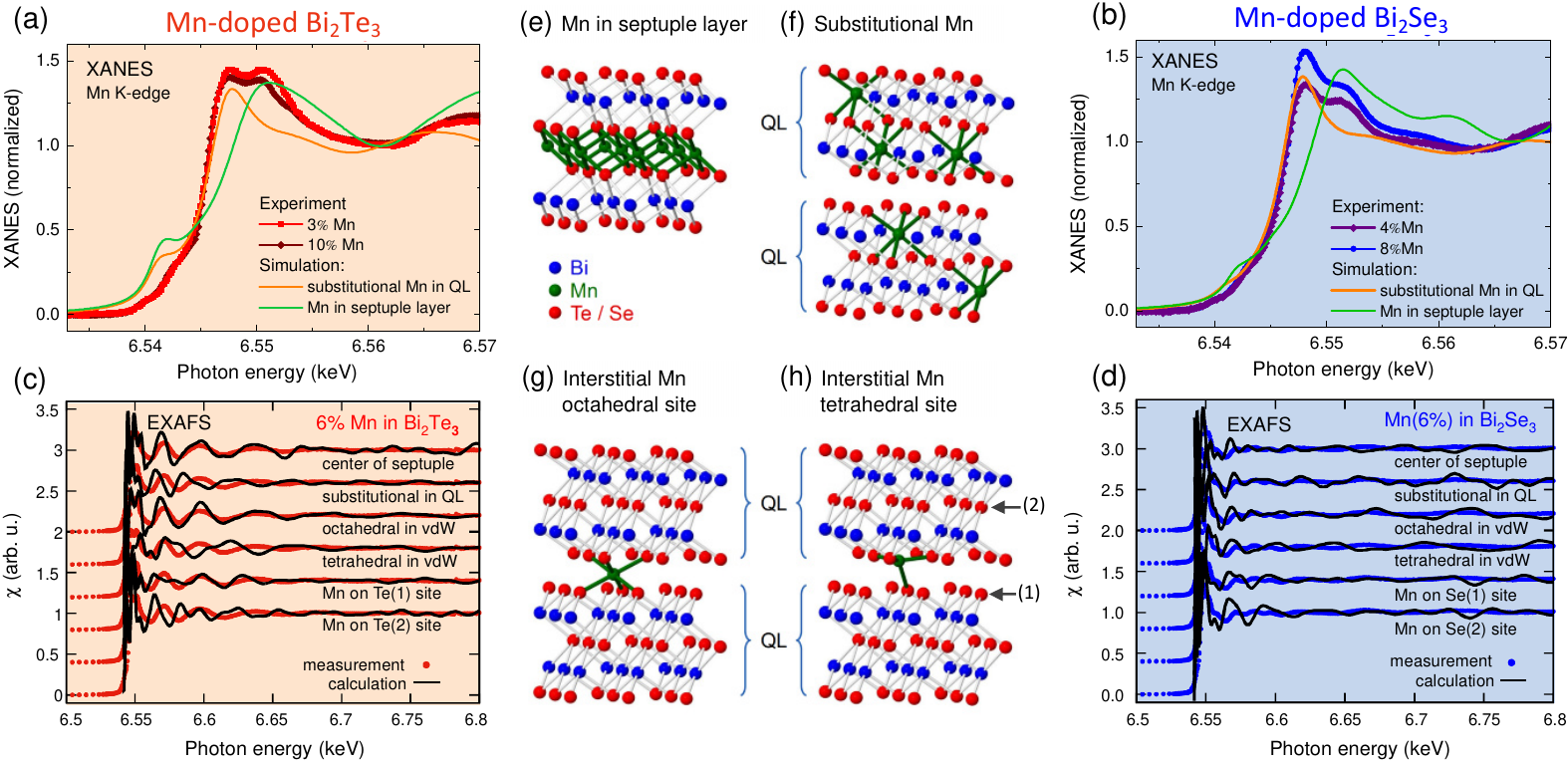}
 	\vspace{1cm}
   \caption{ {\bf{Spectroscopic determination of  the Mn incorporation sites in \BiTe\ and \BiSe\ }}derived from x-ray absorption spectroscopy (XANES, EXAFS) at the Mn K-edge. The symbols represent the experimental spectra, the solid lines the simulation performed for different incorporation sites including substitutional Mn on Bi sites in quintuple layers, Mn in the center of  septuple layers, as well as interstitial Mn in the van der Waals gaps
as shown in (e-h). The two contributions seen in the XANES spectra (a,b) are associated with substitutional Mn sites and Mn in the center of the septuple layers, having a higher weight in \BiTe\ than in \BiSe. The comparison of the EXAFS data (red lines in (c,d)) recorded for 6\%\ Mn to the simulations (green lines) shows best agreement for Mn in the center of the septuple layers in  \BiTe. EXAFS oscillations are less pronounced in \BiSe\ indicating a more pronounced distribution of Mn over the different lattice sites for all Mn concentrations (see Supplementary Information).
}
  \end{center}
\end{figure}

\begin{figure}[t]
  \begin{center}
    \includegraphics[width=\linewidth,keepaspectratio]{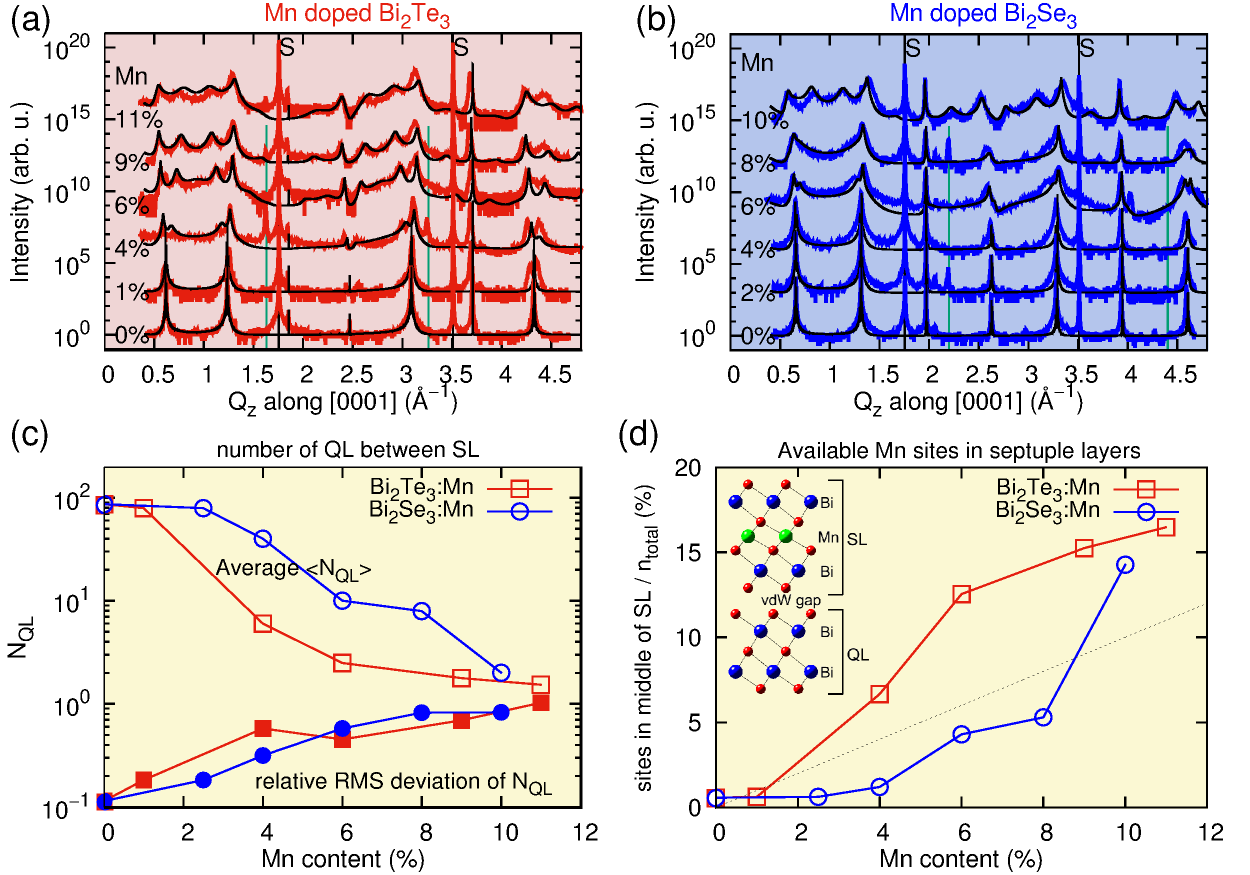}
    \caption{ {\bf{X-ray diffraction analysis of the septuple/quintuple heterostructures}} formed in 
\BiTe\  and \BiSe\ upon Mn doping as a function of Mn concentration ranging from 0 to 11\%. The measured diffraction spectra (red and blue lines in (a,b)) are fitted using a random stacking paracrystal model consisting of a  statistically varying alternation of  Bi$_2X_3$ quintuple and Bi$_2$Mn$X_4$ septuple layers as described in the Supplementary Information, providing an excellent fit (black lines) of the experimental data for both the telluride and selenide system. The average number of quintuples $\left< N_{\rm QL} \right>$  between subsequent septuples and the root mean square (RMS) width of the random distribution derived from the fit is plotted in (c) versus Mn content (open, respectively, full symbols). A smaller average distance $\left< N_{\rm QL} \right>$, i.e., higher concentrations of septuples, is found for \BiTe\  as compared to \BiSe . The number of available Mn sites in the center of the septuple layers relative to the total number $n_{\rm tot}$ of (Bi and Mn) atoms is shown in (d) versus nominal Mn content. The number expected for unity occupancy is indicated by the dashed line. Experimental points below the line indicate that a significant fraction of Mn atoms resides in other lattice sites. This applies to \BiSe\ but not to \BiTe.
}
  \end{center}
\end{figure}

\end{document}